# Between Fear and Desire, the "Monster" Artificial Intelligence (AI): Analysis through the Lenses of Monster Theory

Ahmed Tlili

Smart Learning Institute of Beijing Normal University, Beijing, China
ahmed.tlili23@yahoo.com

**Abstract:**
With the increasing adoption of Artificial Intelligence (AI) in all fields and daily activities, a heated debate is found about the advantages and challenges of AI and the need for navigating the concerns associated with AI to make the best of it. To contribute to this literature and the ongoing debate related to it, this study draws on the Monster theory to explain the conflicting representation of AI. It suggests that studying monsters in popular culture can provide an in-depth understanding of AI and its monstrous effects. Specifically, this study aims to discuss AI perception and development through the seven theses of Monster theory. The obtained results revealed that, just like monsters, AI is complex in nature, and it should not be studied as a separate entity but rather within a given society or culture. Similarly, readers may perceive and interpret AI differently, just as readers may interpret monsters differently. The relationship between AI and monsters, as depicted in this study, does not seem to be as odd as it might be at first.

## 1. Introduction

Timothy Beal (Beal, 2001) mentioned that the word "monster" derives from the Latin *monstrare* (to show) and *monere* (to warn). According to Cohen (1996), Monster theory explores how monsters reflect and shape the societies that create them. Cohen's work frames monsters as complex symbols that are perceived as "other" or outside the norms of society, embodying societal anxieties, and beliefs, but also desire and aspiration.

Building on this, AI can be seen as a "monster" to fear because of its potential in taking humans' jobs, causing job displacement. The idea of machines replacing humans can provoke deep anxieties about loss of employment, dehumanization of several services, and the erosion of interpersonal relationships. On the other hand, AI as monsters that are desired is seen in the possibilities that AI brings for societies and humanities. This contradictory representation of AI was further depicted by Tlili et al. (2023), who questioned AI as "what if the devil is my guardian angel?" In other research, they also called taming this AI "monster" to prevent its negative impact (Tlili et al., 2024a).

Beyond research, Western movies have also pictured this conflicting side of AI and machines, such as *I, Robot* and *Eagle Eye,* where machines and robots were coexisting with humans and protecting them until they decided to improve and reproduce themselves without human involvement, and their goal is to drive their creators—the humans—to extinction.

Therefore, to better understand AI and how it is depicted as a "monster," the present study draws on the Monster theory (Cohen, 1996), specifically the seven theses presented within it (Cohen, 1996, 2007), to analyze how AI is perceived and the conflictual feelings and opinions towards it. While discussing robots as monsters in nursing, Erikson and Salzmann-Erikson (2016) stated that some representations in popular culture (e.g., monsters in this present study) can further contribute to scholarly knowledge related to a given technology.

## 2. AI through the lenses of Monster Theory

This section analyzes AI through the lenses of the Monster Theory, specifically the seven theses (Cohen, 1996, 2007). Each of these theses is discussed in the following subsequent sections.

### 2.1. The monster's body is a cultural body

Monsters were not born; they are products of specific cultures, embodying the anxieties, desires, and fears of that culture. Through this lens, AI can be understood as a "monster" that embodies various societal values, fears, and aspirations, reflecting and impacting the cultural landscape.

AI often reflects societal fears about loss of control, job displacement, and the erosion of privacy (Yarovenko et al., 2024). For instance, media frequently depicts robots and automation as leading to human extinction. These narratives reveal anxieties about technology outpacing human capabilities and the potential consequences of unregulated technological advancement. In this way, AI acts as a mirror reflecting deep-seated fears regarding autonomy, power structures, and economic instability.

The development and deployment of AI technologies often occur within specific cultural and economic frameworks, which may privilege certain values over others—typically those found in Western societies (Muldoon & Wu, 2023). This leads to a form of technological imperialism, where non-Western communities may feel pressured to adopt AI technologies that align with the interests and standards of more developed nations. As a cultural body, AI can perpetuate existing power dynamics, highlighting issues of representation, equity, and access.

Also, different cultures and societies hold diverse ethical perspectives (Ricoeur, 1973). What one society may deem ethically acceptable, another may reject outright, reflecting differing values and beliefs about technology's role in human life. For instance, approaches to data privacy, surveillance, and automation can significantly vary based on cultural context. This difference invites critical reflection on the ethical frameworks governing AI, shaping how societies navigate the responsibilities that come with developing and deploying AI technologies.

Conversely, AI can also embody cultural aspirations for progress and innovation. The narratives surrounding AI often reflect hope for enhanced learning, improved healthcare, and solutions to complex global challenges such as climate change. In this light, AI is a symbol of humanity's quest for advancement and a better future, showcasing our desire to overcome natural limitations and improve the human condition.

In conclusion, when viewed through Cohen's first thesis, AI emerges as a multifaceted cultural monster, embodying both the fears and aspirations of the societies that create it. This duality encourages critical reflections on how we engage with technology and highlights the importance of understanding the cultural narratives that shape our perceptions of AI. As we explore the complex landscape of AI, it is crucial to go beyond the technological capabilities of AI to also discuss the ethical, social, and cultural implications it entails—acknowledging that, like all monsters, AI cannot be understood in isolation but rather as a product of the cultural bodies that shape it.

## 2.2.The Monster Always Escapes

Monsters defy easy categorization and often elude control. They exist in the spaces between order and chaos, reflecting humanity's struggle to define and contain the unknown. These entities challenge boundaries and norms, showing the fluidity of identities. When we relate this thesis to AI, we can see how AI functions as a "monster" that transcends simple labels and challenges our capacity to fully understand or contain it.

Just like monsters, AI was created between the order of careful thinking at first to develop something that can assist humans and make life accessible and easier, and chaos afterwards, where there is a rush to develop anything and put it out for the public, just for profit and power. Additionally, AI has continually evolved in complexity and capability, making it difficult to limit it within a singular definition (Mustafa et al., 2024). What starts as a narrow AI—designed for specific tasks—can rapidly advance toward more general, human-like intelligence, such as AI agents (Kapoor et al., 2024). This ongoing evolution mirrors the nature of monsters, which resist simple classification. The ambiguity surrounding what constitutes "intelligence" blurs lines between human cognition and machine processing, complicating our understanding and control over AI.

Moreover, AI systems can behave in unpredictable ways, often generating results that were not anticipated by their creators. This unpredictability can be seen as a form of the "monster escaping." For instance, machine learning algorithms may develop solutions or strategies that are effective but puzzling, sometimes leading to unintended consequences that challenge human ethical frameworks (Daneshjou et al., 2021). The inability to fully anticipate AI behavior evokes the fear associated with monsters—both are capable of surprising and reshaping the environments they inhabit. Also, the loss of control over decisions made by machines evokes the image of a monster that surpasses its creator's intentions, prompting questions about accountability, ethics, and responsibility.

Also, AI transcends geographic and cultural boundaries, leading to global implications that are often difficult to manage. As AI systems are developed and deployed internationally, their influences escape national regulations and cultural contexts, creating dilemmas that require nuanced and culturally aware solutions. The interconnectedness of AI makes it both a global resource and a potential source of strife, compounding fears around sovereignty, inequality, and the loss of local identity.

In conclusion, viewing AI through the lens of the second thesis of Monster theory reveals its inherent unpredictability and the challenges of control. AI embodies the chaotic nature of a monster that continually escapes the confines of human understanding and governance. While navigating the potential and risks of AI, acknowledging its "monstrous" characteristics allows us to confront the fears and ethical dilemmas it raises, prompting critical discussions around the relationships between technology, humanity, and the ever-evolving landscape of intelligence. Embracing this complexity encourages a more nuanced engagement with AI, urging us to navigate its potential while remaining vigilant about its capacity to transcend our expectations and controls.

**2.3. The Monster Is the Harbinger of Category Crisis**

Monsters disrupt established categories and classifications, forcing societies to confront their own constructs of identity, morality, and values. Such disruption prompts reflection on what is considered normal versus abnormal, appropriate versus inappropriate. This thesis illuminates how AI blurs traditional distinctions between human and machine, intelligence and algorithm, and even moral agency and mechanical operation.

The rapid development of AI has blurred the lines between human and machine intelligence. As machines become capable of performing tasks that were once exclusively human—such as understanding natural language, generating creative works, or making complex decisions—the criteria for what constitutes "intelligence" are called into question. This challenges our long-held assumptions about cognitive capabilities and raises critical ethical questions: Can a machine truly understand or feel, or is it simply mimicking human behavior? The resulting crisis in categories forces society to rethink what it means to be "intelligent."

AI-generated content—whether in art, music, literature, or other creative domains—triggers a crisis in the understanding of authorship and creativity. If an AI can generate a painting or write a novel, what does this mean for the concept of the "creator"? Can machines hold authorship rights, or should they be regarded merely as tools or instruments of human creativity? This challenge undermines traditional definitions of creativity and challenges the uniqueness of human expression, shaking the foundations of artistic and intellectual property. It brings to memory the monkey selfie case (when a monkey named Naruto took selfies with a wildlife photographer) and concepts of originality (Guadamuz, 2016), authorship (Rosati, 2017), and copyright (Guadamuz, 2018).

As AI systems become more autonomous, they raise questions about moral and ethical agency. In scenarios where AI systems make decisions—such as in self-driving cars, medical diagnoses, or law enforcement—who is held accountable for the consequences of those decisions? This blurs the line between human agency and mechanical operation, leading to a crisis in understanding responsibility. If an AI system causes harm, can we attribute guilt to the technology, its creators, or the operators? This ambiguity complicates legal and ethical frameworks, creating a landscape where traditional categories of accountability no longer suffice (Tlili et al., 2024b).

Moreover, AI can challenge cultural and ethical norms concerning privacy, interpersonal relationships, and authenticity. The advent of AI-enabled communication tools, for instance, reshapes how people connect with one another, raising concerns about the dilution of genuine human interaction. This prompts a crisis of values, as societies must grapple with the implications of technology on human relationships and societal cohesion.

In conclusion, viewing AI through the lens of "The Monster Is the Harbinger of Category Crisis" reveals how it disrupts established classifications and prompts critical reflection on the nature of intelligence, creativity, responsibility, and societal values. The emergence of AI presents unique challenges that force societies to rethink their assumptions and frameworks, particularly as these technologies become integrated into daily life. As we confront the category crises spawned by AI, it is important to engage critically with these challenges, exploring ways to redefine our ethical, legal, and social paradigms to accommodate a future in which human and artificial entities coexist. By embracing the complexities embedded in this crisis, we can better navigate the implications of AI and work toward a more equitable and ethical technological landscape.

## 2.4. The Monster Dwells at the Gates of Difference

Monsters often represent something that is perceived as different or foreign—it could be racial, cultural, or ideological. They can serve as a metaphor for societal fears about the "other" and provoke discussions about power dynamics, exclusion, and marginalization. Monsters challenge our understanding of identity, culture, and normality, bringing to light the tensions that arise from encountering the unfamiliar. When we apply this thesis to AI, we can explore how AI represents various forms of difference and challenges established boundaries in society.

AI exists at the intersection of human and non-human intelligence (i.e., hybrid or collaborative intelligence), challenging the boundaries that have traditionally defined what it means to be human. With advanced capabilities in learning, reasoning, and even emotional responses, AI systems can make it difficult to distinguish between human thought processes and computational methods. This challenge evokes fears and questions about the authenticity of human experience and the value of human qualities such as empathy and understanding. Also, as AI systems increasingly interact with people in ways that mimic emotional understanding—such as therapy bots and companion robots—they create a confusion between human interaction and simulated emotions. For example, early conversational systems, such as Eliza (Weizenbaum, 1966), Parry (Colby, 1975), and Alice (Wallace, 2009), were designed to mimic human behavior in a text-based conversation and passed the Turing Test (Turing, 1950; Shieber, 1994) within a controlled scope. This challenge raises questions about what it means to connect with another being, whether human or machine. The emotional responses elicited by AI challenge the boundaries of communication, companionship, and empathy, drawing attention to deep-seated anxieties surrounding loneliness, isolation, and the nature of relationships in a technologically mediated world.

AI technologies challenge traditional roles in various fields—such as education, medicine, and customer service—capitalizing on efficiency and functionality that may reduce the human touch. For instance, the use of AI-driven chatbots or intelligent tutoring systems (ITSs) can alter the roles of educators and support staff, leading to increased concerns about the future of human relationships within these domains. The fear that AI could replace authentic human interactions exemplifies the anxiety surrounding the "otherness" of technology as it challenges established roles and relationships.

AI systems can both reflect and exacerbate societal biases related to race, gender, and identity. Many AI algorithms are trained on datasets that fail to represent the diverse population they are meant to serve, leading to outputs that reinforce stereotypes and marginalize certain groups (Daneshjou et al., 2021). For instance, facial recognition systems have been shown to have higher error rates for people of color and women, leading to unjust outcomes and perpetuating existing inequalities (Ahn et al., 2022; Carter, 2023). This situation creates a difference that prompts societal discomfort about fairness and justice, revealing the systemic issues that lie at the intersection of technology and identity.

In conclusion, when viewed through the lens of "The Monster Dwells at the Gates of Difference," AI emerges as a complex entity that embodies various forms of "otherness." AI challenges traditional definitions of humanity, identity, and ethical frameworks while revealing the tensions and anxieties that arise from these challenges. As we integrate AI into our lives, there is a need to confront these differences and acknowledge the diverse perspectives that shape our understanding of technology. By doing so, we can navigate the complexities and implications of AI more thoughtfully, fostering an environment that respects cultural values, promotes equity, and encourages responsible technological development.

**2.5. The Monster Stands at the Threshold of Becoming**

Monsters indicate transformation—they are often at the intersection of human and non-human or the natural and supernatural. This transformation represents the potential for change, growth, and the evolution of identity, as monsters embody both the possibility of destruction and creation. This notion can be applied to AI to reveal its role as a transformative force in society, one that holds significant promise while simultaneously presenting profound challenges.

AI technology is rapidly advancing, transitioning from narrow AI systems that perform specific tasks to potentially more general forms of AI that can learn and adapt in ways similar to human intelligence. This evolution raises critical questions about the future of both AI and humanity (Qorbani, 2020), exploring possibilities for collaboration (i.e., collaborative intelligence) as well as competition. The higher cognitive function of AI represents something that is not yet fully defined or understood, evoking both excitement about its potential and fear about its implications.

The integration of AI into various sectors, such as healthcare, education, and industry, is transforming the roles of professionals in those fields. Doctors, educators, and workers might increasingly rely on AI to perform tasks, analyze data, and make predictions, thereby altering their roles and responsibilities. This shift stands at the threshold of becoming a new paradigm in which human roles are redefined, often prompting fears about obsolescence even as it opens opportunities for individuals to focus on more complex, creative, or human-centered aspects of their jobs.

As AI systems evolve, they begin to inhabit moral and ethical dilemmas. For example, autonomous vehicles must make decisions in crisis situations, weighing lives and outcomes in ways that raise significant ethical questions (Qorbani, 2020; Yarovenko et al., 2024). This capacity for decision-making represents a threshold condition where AI transitions from merely executing programmed commands to engaging in moral reasoning. This evolution requires society to grapple with accountability and ethics, examining how we define moral agency when it comes to machines. Also, when integrating AI into predictive models—such as in finance, healthcare, and social dynamics—there is a tension between the desire for greater predictability and the inherent uncertainty of human behavior and society. AI has the capability to analyze vast datasets and generate insights that seem to promise control over our environments, but these systems also reveal the unpredictability of human action.

In conclusion, when analyzed through the lens of "The Monster Stands at the Threshold of Becoming," AI emerges as a transformative force that embodies both potential and risk. With the rapid development of AI, it is crucial to rethink who we are and what we can become in the face of this change. Simply remaining as the "slaves" of AI, by feeding it data and enhancing its cognitive capabilities (processors), without any focus on human development and empowerment might lead to AI overtaking humans. AI must empower humans and complement them rather than replace them. Recognizing AI as a monster can help us to understand its societal implications and encourages proactive engagement accordingly. This threshold moment calls for active involvement in the development of AI governance, regulatory measures, and societal norms to ensure that AI helps create a more equitable future with a balanced hybrid intelligence (human intelligence and AI).

**2.6. Fear of the Monster Is Really a Fear of the Self**
The monster can reflect internal anxieties about ourselves, representing aspects of our identity, desires, or impulses that we reject or suppress. Instead of merely being a fear of the "other," the monster often symbolizes a confrontation with our own darker tendencies and vulnerabilities. When we apply this thesis to AI, we can explore how our fears surrounding AI often reveal deeper concerns about ourselves as individuals and as a society.

A key concern about AI is that many people fear they lack the skills to use it effectively, which fosters feelings of incompetence and drives strong resistance to adopting it in their contexts (Su et al., 2023). To address this issue, several organizations, such as UNESCO, have proposed some AI literacy frameworks for students (Miao & Shiohira, 2024) and teachers (Cukurova & Miao, 2024) to increase AI adoption in education. The fear of AI could also stem from societal pressure. For example, a recent study in China revealed that students are choosing to learn AI mostly out of 'guilt or shame,' not out of personal motivation (Li et al., 2025).

Also, AI systems are often trained on data that reflects human behaviors, thoughts, and biases. Thus, when we encounter biases in AI—such as in algorithms that perpetuate racial or gender discrimination—we are confronted with our own societal flaws and prejudices (Tlili & Burgos, 2025). The imperfections of AI serve as a mirror, urging us to reflect on the disparities and injustices that exist within our communities. Our fear of these biases can be an expression of discomfort with our collective moral failings and an indication of the work that lies ahead in confronting and addressing these inequities.

The fear is also associated with over-relying on or losing control over technology that has the potential to surpass human intelligence. This fear can be interpreted as a reflection of our human anxieties regarding independency, autonomy, and agency. As we create machines capable of independent thought and decision-making, we may confront unsettling questions about our place in the world and the reliability of our judgment. This fear of AI mirrors our deeper concern about coping with the complexities of a rapidly changing landscape in which we might feel increasingly powerless. On the other hand, our increasing reliance on AI for everyday tasks and decision-making can evoke fears of dependency (Zhai et al., 2024). This reliance serves to highlight our vulnerabilities as humans in an age of automation and rapid technological advancement. The fear of becoming too dependent on AI for critical thinking or social interaction indicates a discomfort with our human limitations and weaknesses. It compels us to explore what it means to remain independent, authentic, and capable in a world where technology plays an increasingly central role.

The rise of AI-generated content—whether in art, literature, or social media—elicits concerns about authenticity and identity (Tlili et al., 2025). As machines produce creative works or simulate human interactions, we may become anxious about the nature of human creativity and originality. The fear that AI can replicate or even enhance human expression raises questions about what it means to be genuinely human. It often reveals deeper insecurities about individual values and contributions, leading us to grapple with vulnerabilities around identity and the essence of our humanity.

The apocalyptic narratives surrounding AI—where machines turn against humanity or lead to cataclysmic outcomes—can signify a deeper existential fear about the mortality and the trajectory of human civilization. These narratives serve as cautionary tales that compel us to examine our choices, ethical frameworks, and relationship with technology. The fear of a rogue AI often reflects a fear of our own destructive capabilities, urging us to confront the moral complexities of innovation and the potential consequences of our actions.

In conclusion, when analyzed through the lens of "Fear of the Monster Is Really a Fear of the Self," AI emerges as a profound reflection of our inner struggles, anxieties, and insecurities. The fears we harbor about AI often stem from deeper concerns about our identity, agency, morality, and how society might look at us. By examining these fears, we have the opportunity to engage in critical self-reflection, encouraging a more nuanced reflection of our relationship with technology. Acknowledging the ways in which AI reflects and amplifies human fears can empower us to cultivate resilience, responsibility, and ethical engagement in the development and application of AI technologies. Ultimately, this awareness invites us to confront our own identities, values, and aspirations in our pursuit of a positive future framed by AI.

## 2.7. The Monster Carries the Sign of Our Times

The seventh thesis of Monster theory, "The Monster Carries the Sign of Our Times," asserts that monsters embody the concerns, beliefs, and tensions of the society that creates and encounters them. They are reflections of the social, political, and cultural landscapes of their era, serving as barometers for societal challenges and aspirations. When applying this thesis to AI, we can see how it encapsulates the anxieties, hopes, and ethical dilemmas present in contemporary society, marking it as a quintessential monster of our times.

The deployment of AI technologies often creates social inequalities because of the uneven distribution in accessing technologies worldwide (Zajko, 2021). This disparity is reflective of the growing wealth gap and socio-economic divides in society. As AI systems can perpetuate and amplify biases, their presence underscores the urgent need to address systemic injustices. On the other hand, AI has the potential to address global challenges, such as climate change (Cheong et al., 2022), highlighting its role as a "sign" of our times. By developing AI systems that can optimize energy use, improve resource management, and promote sustainable practices, we leverage technology to combat pressing existential threats. This intersection of AI with ecological concerns speaks to a growing awareness of interdependence between humanity and the environment, illustrating how current societal crises influence technological development.

Concurrent with fears, AI also embodies hope for progress, promising advancements in fields like healthcare, education, and environmental sustainability. The potential for AI to solve complex problems and improve human lives reflects the human aspiration for innovation and improvement. This duality captures the zeitgeist of our times: while we grapple with the ethical dilemmas posed by AI, we also cling to the belief that technology can enhance our existence and contribute to societal advancement.

In conclusion, when viewed through the lens of "The Monster Carries the Sign of Our Times," AI emerges not only as a powerful technological force but also as a complex social symbol intertwined with contemporary anxieties, hopes, and ethical dilemmas. AI carries with it the weight of our cultural narratives and societal challenges, reflecting what we value, fear, and aspire to as a civilization. By engaging critically with the roles and implications of AI in our lives, we can better navigate the complexities of the present moment while shaping a future that aligns with our shared values and collective aspirations. Understanding AI as both a product and a reflection of our times allows us to actively participate in the ethical and social discourse surrounding its development, ensuring that it serves as a tool for equity, empowerment, and positive transformation in society.

## 3. Conclusions and final remarks

This study analyzes AI through the lenses of Monster theory. By recognizing AI as a multifaceted "monster" that carries the signs of our times, we can better navigate the complexities and challenges posed by this technology.

The findings suggest that the rapid AI development has catalyzed serious debate about adopting in reality the fictional Three Laws of Robotics, proposed by Asimov (1942) in his story *Runaround*. These laws are: (1) robots must not harm humans or allow harm through inaction, (2) robots must obey human orders unless they conflict with the First Law, and (3) robots must protect their own existence unless it violates the First or Second Law. The study further argues that AI should not be viewed merely as a standalone technology but as a complex ecosystem of interconnected variables. Embracing a holistic, ethical, and inclusive approach to AI design and development can allow realizing its potential while addressing the fears and concerns it evokes. Ultimately, the goal should be to create a future in which AI serves as an ally that enhances human life, promotes equity, and fosters communal growth, ensuring that technology reflects our highest values and aspirations.

Finally, it is important to highlight that the way monsters are represented and the values they carry are based on writers' views and thoughts; however, these views and thoughts are not necessarily those of readers and might also be interpreted differently by them. When applying this to AI, it is found that several AI-based systems (e.g., generative AI or large language models) were developed, by developers and companies, and pushed to the public as a tool that can support in given tasks, but these tools backfired, raising many concerns. Also, in some instances, some of these AI-based tools were perceived differently in some specific contexts or cultures.

Cohen, J. J. (Ed.). (1996). Monster theory [electronic resource]: reading culture. U of Minnesota Press.

Colby, K. (1975). *Artificial Paranoia: A Computer Simulation of Paranoid Processes.* Elsevier.

Cukurova, M., & Miao, F. (2024). AI competency framework for teachers. UNESCO Publishing.

Daneshjou, R., Smith, M. P., Sun, M. D., Rotemberg, V., & Zou, J. (2021). Lack of transparency and potential bias in artificial intelligence data sets and algorithms: a scoping review. *JAMA dermatology*, 157(11), 1362-1369. https://doi.org/10.1001/jamadermatol.2021.3129

Erikson, H., & Salzmann-Erikson, M. (2016). Future challenges of robotics and artificial intelligence in nursing: what can we learn from monsters in popular culture? *The Permanente Journal*, 20(3), 15-243. http://dx.doi.org/10.7812/TPP/15-243.

Guadamuz, A. (2018). Can the monkey selfie case teach us anything about copyright law? *WIPO Magazine*, 1, 40–46.

Guadamuz, A. (2016). The monkey selfie: Copyright lessons for originality in photographs and internet jurisdiction. *Internet Policy Review*. https://doi.org/10.14763/2016.1.398

Kapoor, S., Stroebl, B., Siegel, Z. S., Nadgir, N., & Narayanan, A. (2024). Ai agents that matter. *arXiv preprint arXiv*:2407.01502.

Li, J., Zhang, J., Chai, C. S., Lee, V. W., Zhai, X., Wang, X., & King, R. B. (2025). Analyzing the network structure of students' motivation to learn AI: a self-determination theory perspective. *npj Science of Learning*, 10(1), 48. https://doi.org/10.1038/s41539-025-00339-w

Miao, F., & Shiohira, K. (2024). AI competency framework for students. UNESCO Publishing.

Muldoon, J., & Wu, B. A. (2023). Artificial intelligence in the colonial matrix of power. *Philosophy & Technology*, 36(4), 80. https://doi.org/10.1007/s13347-023-00687-8

Mustafa, M. Y., Tlili, A., Lampropoulos, G., Huang, R., Jandrić, P., Zhao, J., ... & Saqr, M. (2024). A systematic review of literature reviews on artificial intelligence in education (AIED): a roadmap to a future research agenda. *Smart Learning Environments*, 11(1), 59. https://doi.org/10.1186/s40561-024-00350-5